\documentclass[12pt]{article}
\usepackage{graphicx}
\newcommand{\bfb}{\cal}

\begin{document}

{\bf\Large Giant magnetoresistance in  semiconductor / granular film
heterostructures with cobalt nanoparticles } \\{ }\\ {L.V. Lutsev${
}^1$, A.I. Stognij${ }^2$, and N.N. Novitskii${ }^2$} {\\{ }\\ \it
${ }^1$A.F. Ioffe Physico-Technical Institute, Russian Academy of
Sciences, 194021 St. Petersburg, Russia \\ ${
}^2$Scientific-Practical Materials Research Center of National Academy of Sciences of Belarus, 220072 Minsk, Belarus} {\\
l\_lutsev@mail.ru}

\begin{abstract}
We have studied the electron transport in SiO${ }_2$(Co)/GaAs and
SiO${ }_2$(Co)/Si heterostructures, where the SiO${ }_2$(Co)
structure is the granular SiO${ }_2$ film with Co nanoparticles. In
SiO${ }_2$(Co)/GaAs heterostructures giant magnetoresistance effect
is observed. The effect has positive values, is expressed, when
electrons are injected from the granular film into the GaAs
semiconductor, and has the temperature-peak type character. The
temperature location of the effect depends on the Co concentration
and can be shifted by the applied electrical field. For the SiO${
}_2$(Co)/GaAs heterostructure with 71 at.\% Co the magnetoresistance
reaches 1000 ($10^5$ \%) at room temperature. On the contrary, for
SiO${ }_2$(Co)/Si heterostructures magnetoresistance values are very
small ($4 \%$) and for SiO${ }_2$(Co) films the magnetoresistance
has an opposite value. High values of the magnetoresistance effect
in SiO${ }_2$(Co)/GaAs heterostructures have been explained by
magnetic-field-controlled process of impact ionization in the
vicinity of the spin-dependent potential barrier formed in the
semiconductor near the interface. Kinetic energy of electrons, which
pass through the barrier and trigger the avalanche process, is
reduced by the applied magnetic field. This electron energy
suppression postpones the onset of the impact ionization to higher
electric fields and results in the giant magnetoresistance. The
spin-dependent potential barrier is due to the exchange interaction
between electrons in the accumulation electron layer in the
semiconductor and $d$-electrons of Co. Existence of spin-polarized
localized electron states in the accumulation layer results in the
temperature-peak type character of the barrier and the
magnetoresistance effect. Spin injector and spin-valve structure on
the base of ferromagnet / semiconductor heterostructures with
quantum wells with spin-polarized localized electrons in the
semiconductor at the interface are considered.
\end{abstract}

\section{Introduction}
Electron spin transport in ferromagnet / semiconductor (FM / SC)
heterostructures has recently become an active area of research. The
manipulation of carrier spin in FM / SC heterostructures offers
enhanced functionality of spin-electronic devices such as spin
transistors, sensors and magnetic memory cells~\cite{ref1,ref2}. FM
/ SC heterostructures are intended to employ as magnetoresistance
cells and injectors of spin-polarized electrons in
SCs~\cite{ref3,ref4,ref5}. For practical applications it is highly
desirable to realize these effects at room temperature. Spin
transport phenomena and magnetoresistance are observed on a number
of heterostructures.

\noindent 1. Spin injection into a non-magnetic SC is observed at
low temperatures in magnetic SC / non-magnetic SC heterostructures
\cite{ref6,ref7,ref8} and in ferromagnetic metal / non-magnetic
SC~\cite{ref9,ref10,ref11,ref12,ref13,ref14}. At room temperature
the spin injection reveals low efficiency.

\noindent 2. Spin injection in the ferromagnetic metal / insulator /
SC heterostructure is more efficient in comparison with the spin
injection from ferromagnetic metal / SC
heterostructures~\cite{ref15,ref16,ref17,ref18,ref19}. The maximum
of the spin polarization of injected electrons is achieved for a MgO
barrier on GaAs (47~\% at 290~K)~\cite{ref17}.

\noindent 3. The giant magnetoresistance (GMR) is observed in metal
magnetic multilayers~\cite{ref20,ref21,ref22,ref23}. For three-layer
structures, the typical values of GMR at room temperature lie in the
range 5 - 8~\%.

\noindent 4. High values of tunneling magnetoresistance (TMR) are
realized on the base of magnetic tunnel junction (MTJ)
structures~\cite{ref24,ref25,ref26,ref27,ref28,ref29,ref30,ref31,ref32,ref33,ref34}.
Spin-dependent tunneling is not only determined by the properties of
ferromagnetic electrods but also depends on the electronic structure
of insulator barriers. The maximum TMR ratio of 500~\% at room
temperature was observed in the MTJ structure with the MgO
barrier~\cite{ref34}.

\noindent 5. Extremely large magnetoresistance can be achieved by
use magnetic-field-dependent avalanche breakdown
phenomena~\cite{ref35,ref36,ref37,ref38,ref39,ref40,ref41,ref42,ref43}.
Values of the magnetoresistance effect based on the avalanche
breakdown reach $10^5$ \% in the Au / semi-insulating GaAs Schottky
diode at room temperature~\cite{ref37}.

Although important results in the spin injection and in the
magnetoresistance have been obtained, the efficient spin injection
at room temperature has not been achieved and for some applications
it is necessary to use sensors with high magnetoresistance values.
These problems can be resolved by using FM / SC heterostructures
with spin-dependent potential barrier, which governs the kinetic
energy of injected electrons and the onset of impact
ionization~\cite{ref38,ref39,ref40}. In these heterostructures FM is
a granular film with $d$ (or $f$) metal nanoparticles. In contrast
with metal / SC structures with the Schottky barrier based on the
magnetic-field-dependent avalanche breakdown
phenomena~\cite{ref37,ref41,ref42,ref43}, the transparency of the
spin-dependent potential barrier, which is formed in the
spin-polarized accumulation electron layer in the SC near the
interface, is characterized by the temperature-peak dependence and
is differ for different spin orientations of injected electrons. The
barrier is due to the exchange interaction between $d$ ($f$)
electrons in the FM at the interface and electrons in the SC, which
polarizes electrons in the accumulation layer.

In this paper, we study the magnetoresistance in SiO${ }_2$(Co)/GaAs
and SiO${ }_2$(Co)/Si heterostructures, where the SiO${ }_2$(Co) is
the granular SiO${ }_2$ film with Co nanoparticles. Sample
preparation and experimental results are presented in section 2. The
effect is more expressed, when electrons are injected from the
granular film into the SC, therefore, the magnetoresistance has been
called the injection magnetoresistance (IMR)~\cite{ref38,ref39}. For
SiO${ }_2$(Co)/GaAs heterostructures the IMR value reaches 1000
($10^5$ \%) at room temperature, which is two-three orders higher
than maximum values of the GMR in metal magnetic multilayers and the
TMR in MTJ structures. On the contrary, for SiO${ }_2$(Co)/Si
heterostructures the magnetoresistance values are very small and for
SiO${ }_2$(Co) films the intrinsic magnetoresistance is of a
negative value. The IMR effect has a temperature-peak type character
and its location can be shifted by the applied electrical field.
High values of the IMR effect in SiO${ }_2$(Co)/GaAs
heterostructures and the temperature-peak type character are
explained in section 3 by the theoretical model of a
magnetic-field-controlled avalanche process provided by electrons
passed through the spin-dependent potential barrier in the
accumulation layer at the interface~\cite{ref40}. In section 4 we
consider FM / SC heterostructures with quantum wells with
spin-polarized localized electrons in the SC at the interface as
efficient room-temperature spin injectors and magnetic sensors.
These heterostructures can be used as bioanalytical sensors with
higher sensitivity in comparison with GMR-sensors~\cite{ref44,ref45}
and as injectors in spin-valve transistors and in spin field-effect
transistor (FET) structures~\cite{ref46,ref47,ref48}.

\section{Experimental results}
\subsection{Sample preparation}

Experiments were performed on samples of amorphous silicon dioxide
films containing cobalt nanoparticles grown (1) on gallium arsenide,
(SiO${ }_2$)${ }_{100-x}$Co${ }_x$/GaAs (or shorter SiO${
}_2$(Co)/GaAs), (2) on silicon, (SiO${ }_2$)${ }_{100-x}$Co${
}_x$/Si (or shorter SiO${ }_2$(Co)/Si), and (3) on quartz
substrates. n-GaAs substrates with thickness of 0.4~mm are of the
(100)-orientation type. Electrical resistivity of GaAs chips was
measured by the dc four-probe method at room temperature and was
equal to 0.93$\cdot 10^5$~$\Omega\cdot$cm. The 0.4 mm n-Si
substrates have the orientation of (100) and the resistivity of
3.7~$\Omega\cdot$cm. Prior to the deposition process, substrates
were polished by a low-energy oxygen ion beam~\cite{ref49,ref50}.
The roughness height of the polished surfaces did not exceed 0.5 nm.

\begin{figure*}
\begin{center}
\includegraphics*[scale=0.8]{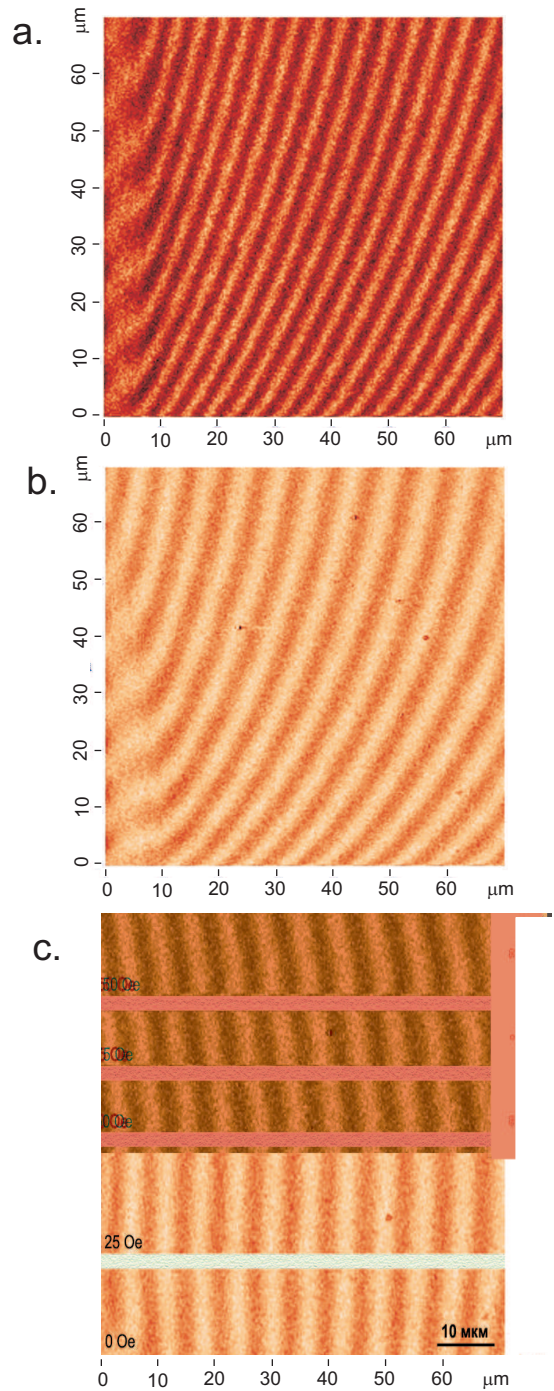}
\end{center}
\caption{Magnetic field microscope image of the domain structure on
samples with SiO${ }_2$(Co) films with 82 at.\% Co (a) on the GaAs
substrate and (b) on the Si substrate. (c) Influence of the applied
magnetic field on the domain structure on the SiO${ }_2$(Co)/Si
sample with 82 at.\% Co.} \label{Fig1}
\end{figure*}

The SiO${ }_2$(Co) films were deposited by ion-beam co-sputtering of
the composite cobalt-quartz target onto GaAs, Si and quartz
substrates heated to 200$^{\circ}$C. The concentration of Co
nanoparticles in the silicon dioxide deposit was varied by changing
the ratio of cobalt and quartz target areas. The film composition
was determined by the nuclear physical methods of element analysis
using a deuteron beam of the electrostatic accelerator (PNPI,
Gatchina, Leningrad region, Russia). The cobalt to silicon atomic
ratio was measured by the Rutherford backscattering spectrometry of
deuterons. The oxygen concentration in films was determined by the
method of nuclear reaction with deuterons at $E_d$ = 0.9 MeV: ${
}^{16}$O $+d\to p+{ }^{17}$O. This technique is described in more
detail elsewhere~\cite{ref51}. For the samples studied, the relative
content of cobalt $x$ and the film thickness are listed in Table 1.
The average size of Co particles was determined by the small-angle
X-ray scattering and increased as the concentration of $x$ grows:
from 2.7~nm at $x$ = 38 at.\% to 4.4~nm at $x$ = 82 at.\%. Cobalt
particles are in the ferromagnetic state~\cite{ref52,ref53,ref54}.
The samples with high concentration of Co (71 and 82 at.\%) exhibit
ferromagnetic behaviour confirmed by the presence of a domain
structure (Figure \ref{Fig1}) obtained with NT-MDT magnetic field
microscope Solver HV-MFM. The period of the domain structure for the
SiO${ }_2$(Co)/GaAs sample with 82 at.\% Co is equal to 3.9 $\mu$m,
which is smaller than the domain period for the same SiO${ }_2$(Co)
film on the Si substrate (6.0 $\mu$m). The samples with low
concentration of Co are superparamagnetic.

Electrical resistivity of SiO${ }_2$(Co) films was measured by the
dc four-probe method on SiO${ }_2$(Co)/ quartz heterostructures at
room temperature. As the Co content increased, the resistivity of
SiO${ }_2$(Co) films decreased from 1.46$\cdot 10^2$~$\Omega\cdot$cm
(38 at.\%) to 1.1~$\Omega\cdot$cm (82 at.\%).\\{ }\\
{\bf Table 1.} Properties of SiO${ }_2$(Co) films sputtered on GaAs,
Si and quartz substrates.\\{ }\\

\begin{tabular}{c|ccc}
\hline\hline
Co concentration & \multicolumn{3}{|c}{Film thickness (nm)}\\
\cline{2-4}
$x$ (at.\%) & GaAs substrate & Si substrate & Quartz substrate\\
\hline
38 & 86 & 86 & 860\\
45 & 81 & 81 & 810\\
54 & 90 & 90 & 900\\
71 & 95 & 95 & 950\\
82 & 95 & 95 & 950\\
\hline\hline
\end{tabular}\\{ }\\

\subsection{Experiment}

\begin{figure*}
\begin{center}
\includegraphics*[scale=.45]{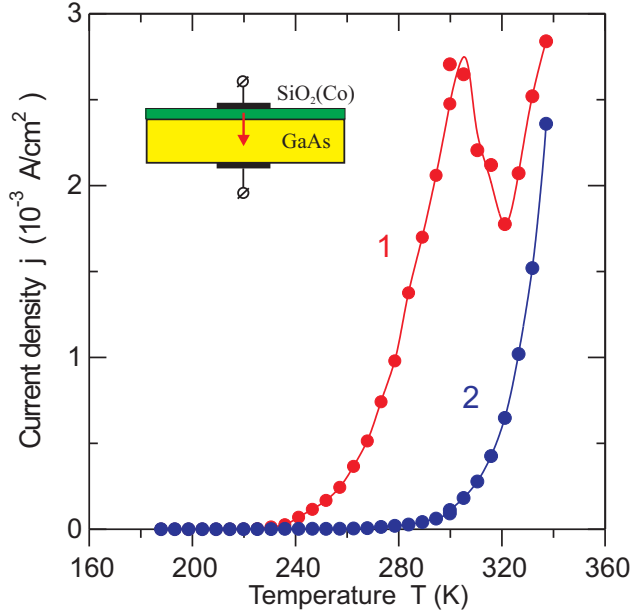}
\end{center}
\caption{Temperature dependencies of the inject current $j$ for the
SiO${ }_2$(Co)/GaAs structure with the Co concentration 71~at.\% at
the applied voltage $U$ = 70~V. (1) In the absence of a magnetic
field, (2) in the magnetic field $H$ = 10~kOe. $H$ is parallel to
the surface of the SiO${ }_2$(Co) film. Solid lines are guides for
the eye.} \label{Fig2}
\end{figure*}

We have studied the electron transport and magnetoresistance in
SiO${ }_2$(Co)/SC heterostructures (Table 1). One contact was on the
semiconductor substrate, and the other -- on the SiO${ }_2$(Co)
granular film. All SiO${ }_2$(Co)/GaAs samples and SiO${ }_2$(Co)/Si
samples with the Co concentration, which is equal to or lesser than
71~at.\%, have current-voltage dependencies of the diode type. At
positive voltages for structures of the diode type current-voltage
characteristic electrons are injected from the granular film into
the SC and the current density $j$ is high. For the applied voltage
$U$ = 90~V the current density reaches 6.0$\cdot 10^{-2}$~A/cm${
}^2$. In the case when the applied voltage $U$ is negative,
electrons drift from the SC into the granular film and the current
density is low. For SiO${ }_2$(Co)/Si heterostructure with high Co
content (82 at.\%), the current-voltage characteristic is close to
the dependence of the Ohm type. Figure \ref{Fig2} shows temperature
dependencies of the electron inject current density $j$ for the
SiO${ }_2$(Co)/GaAs structure with the Co concentration $x$ =
71~at.\% at the applied voltage $U$ = 70~V. The resistivity of GaAs
is higher than the resistivity of the film and the applied voltage
primarily falls on the SC substrate. We notice that at the
temperature $T$ = 320~K in the absence of a magnetic field the
inject current has local minimum. The electron inject current
flowing from the granular film into the SC is suppressed by the
magnetic field. The magnetic field $H$ is equal to 10~kOe and is
parallel to the surface plane of the granular film. At $T >$ 320~K
temperature dependencies of the inject current in the absence of a
magnetic field and in the field $H$ are close.

\begin{figure*}
\begin{center}
\includegraphics*[scale=.45]{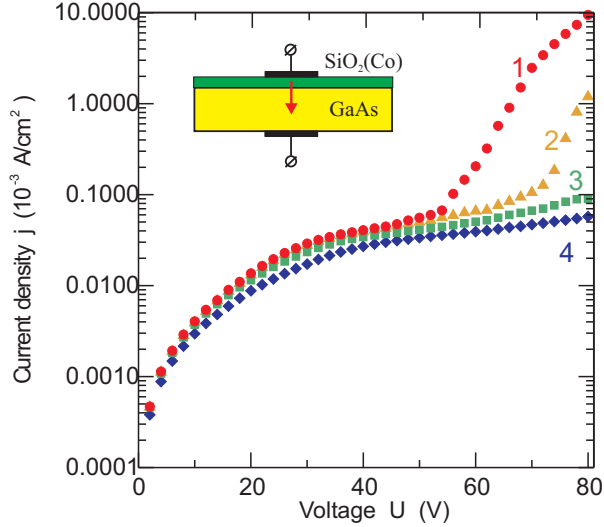}
\end{center}
\caption{Current-voltage characteristic for the injection of
electrons into the semiconductor for the SiO${ }_2$(Co)/GaAs
structure with 71~at.\% Co at different magnetic fields: (1) $H$ =
0, (2) 5~kOe, (3) 10~kOe, (4) 15~kOe. $H$ is parallel to the surface
of the SiO${ }_2$(Co) film.} \label{Fig3}
\end{figure*}

Figure \ref{Fig3} illustrates the effect of the magnetic field on
the current-voltage characteristic for the injection of electrons
into the semiconductor for the SiO${ }_2$(Co)/GaAs structure with
71~at.\% Co. For $U >$ 52~V, a sharp increase in current due to the
process of impact ionization is observed. The applied magnetic field
postpones this process to higher electric fields. The magnetic field
$H$ is parallel to the film surface. If the magnetic field is
perpendicular to the film surface, the dependence of the current on
the magnetic field $H$ is weaker because of the demagnetization
factor of the film, but the magnetic suppression of the current is
still observed.

\begin{figure*}
\begin{center}
\includegraphics*[scale=.45]{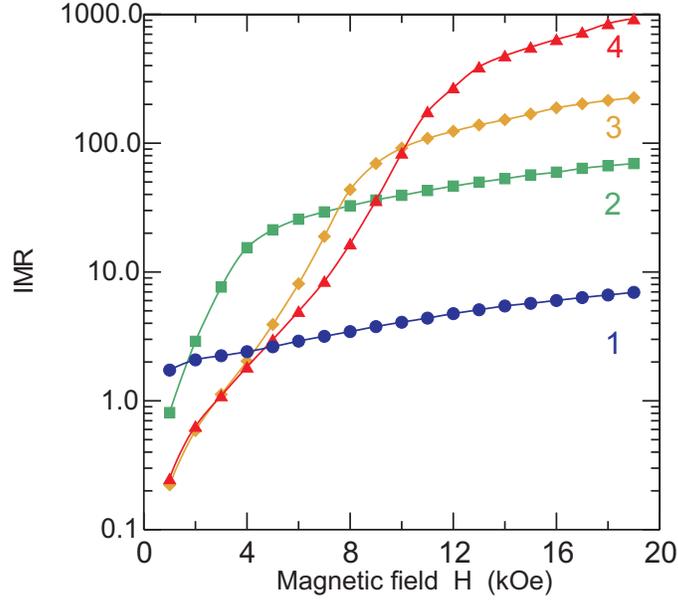}
\end{center}
\caption{Injection magnetoresistance ratio, IMR, versus the magnetic
field $H$ at room temperature for the SiO${ }_2$(Co)/GaAs structure
with 71~at.\% Co at applied voltages: (1) $U$ = 60~V, (2) 70~V, (3)
80~V, (4) 90~V. $H$ is parallel to the surface of the SiO${ }_2$(Co)
film. Solid lines serve to guide the eye.} \label{Fig4}
\end{figure*}

By analogy with GMR and TMR
coefficients~\cite{ref20,ref21,ref22,ref23,ref24,ref25,ref26,ref27,ref28,ref29,ref30,ref31,ref32,ref33,ref34},
we define the injection magnetoresistance coefficient {\it IMR} as
the ratio~\cite{ref38,ref39,ref40}

\begin{equation}
{\mbox{\it IMR}}= \frac{R(H)-R(0)}{R(0)}= \frac{j(0)-j(H)}{j(H)},
\label{eq1}
\end{equation}

\noindent where $R(0)$ and $R(H)$ are the resistances of the SiO${
}_2$(Co)/SC heterostructure without a field and in the magnetic
field $H$, respectively; $j(0)$ and $j(H)$ are the current densities
flowing in the heterostructure in the absence of a magnetic field
and in the field $H$. The IMR ratio for the SiO${ }_2$(Co)/GaAs
structure with 71~at.\% Co at different applied voltages at room
temperature (21$^{\circ}$C) is shown in Figure \ref{Fig4} as a
function of the magnetic field $H$ parallel to the film. As seen
from Figure \ref{Fig4}, the IMR coefficient increases with the
growth of the applied voltage. At the voltage $U$ = 90~V for this
structure the value of IMR reaches up to 1000 ($10^5$ \%) at room
temperature at the field $H$ = 19~kOe. This is two-three orders
higher than maximum values of GMR in metal magnetic multilayers and
TMR in MTJ structures.

\begin{figure*}
\begin{center}
\includegraphics*[scale=.45]{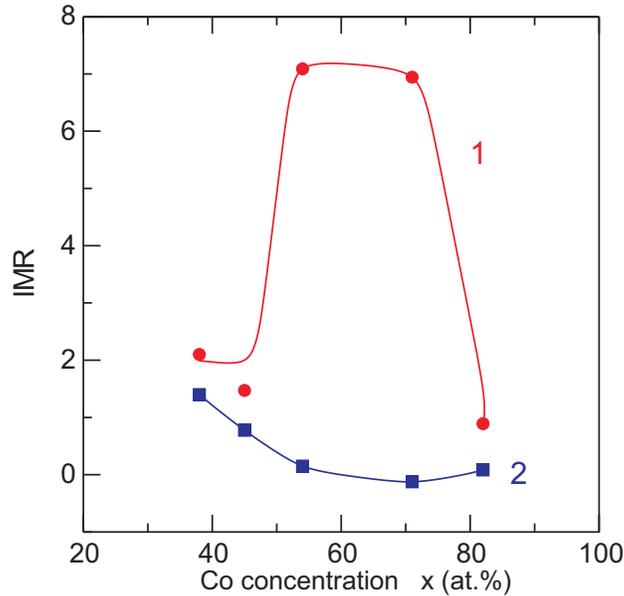}
\end{center}
\caption{Magnetoresistance ratio, IMR, versus the Co concentration
$x$ for SiO${ }_2$(Co)/GaAs structures in the field $H$ = 20~kOe at
the applied voltage $U$ = 60~V for different current directions. (1)
Electrons are injected from the SiO${ }_2$(Co) film into GaAs, (2)
electrons drift from GaAs into the granular film. $H$ is parallel to
the surface of the SiO${ }_2$(Co) film. Solid lines serve to guide
the eye.} \label{Fig5}
\end{figure*}

The IMR ratio for SiO${ }_2$(Co)/GaAs structures versus the Co
concentration $x$ in the in-plane field $H$ = 20~kOe at the applied
voltage $U$ = 60~V for different current directions is presented in
Figure \ref{Fig5}. The IMR coefficient has maximum values for
structures with Co concentrations in the range [54 - 71~at.\%], when
electrons are injected from the SiO${ }_2$(Co) film into the SC. The
IMR ratio decreases for structures with higher ($x >$ 71~at.\%) and
lower ($x <$ 54~at.\%) Co concentrations. On the contrary, in the
case of the opposite current direction (electrons drift from the SC
into the granular film) the magnetoresistance effect becomes less
expressed.

\begin{figure*}
\begin{center}
\includegraphics*[scale=.45]{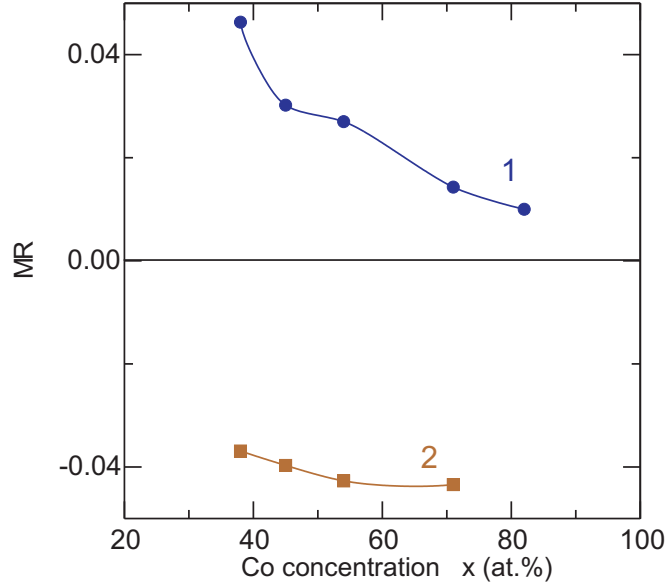}
\end{center}
\caption{Magnetoresistance ratio, MR, versus the Co concentration
$x$ for (1) SiO${ }_2$(Co)/Si structures and for (2) SiO${ }_2$(Co)
films in the in-plane magnetic field $H$ = 20~kOe. Solid lines serve
to guide the eye.} \label{Fig6}
\end{figure*}

As we can see from Figures \ref{Fig4} and \ref{Fig5}, for SiO${
}_2$(Co)/GaAs structures the IMR coefficient can reach high values
at room temperature. In contrast with this, for SiO${ }_2$(Co)/Si
structures magnetoresistance values are very small and the intrinsic
magnetoresistance of SiO${ }_2$(Co) films has negative values
(Figure \ref{Fig6}). The magnetoresistance ratio (MR) for SiO${
}_2$(Co) films is determined by the relation analogous to Eq.
(\ref{eq1}). For SiO${ }_2$(Co)/Si structures electrons are injected
from the granular film into the Si substrate. Taking into account
low values of the resistivity of Si substrates, experiments were
carried out at the applied voltage $U$ = 3~V. For SiO${ }_2$(Co)
films the intrinsic magnetoresistance ratio was measured by the dc
four-probe method on SiO${ }_2$(Co)/quartz samples in the
current-in-plane geometry at the applied voltage $U$ = 60~V at room
temperature.

\begin{figure*}
\begin{center}
\includegraphics*[scale=.45]{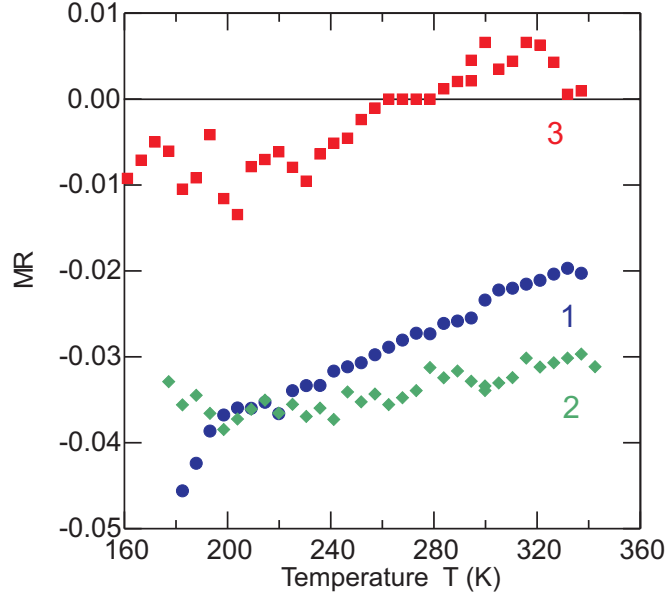}
\end{center}
\caption{Temperature dependencies of the magnetoresistance, MR, for
SiO${ }_2$(Co) films (1) with $x$ = 38~at.\% Co, (2) with $x$ =
71~at.\% Co, and (3) for the SiO${ }_2$(Co)/Si structure with $x$ =
71~at.\% Co content in the in-plane magnetic field $H$ = 10~kOe. }
\label{Fig7}
\end{figure*}

Temperature dependencies of the magnetoresistance can give useful
information about the nature of the magnetoresistance effect. Figure
\ref{Fig7} presents temperature dependencies of the intrinsic
magnetoresistance for SiO${ }_2$(Co) films with low ($x$ = 38~at.\%)
and high ($x$ = 71~at.\%) Co concentrations and for the SiO${
}_2$(Co, 71~at.\%)/Si structure. Experiments were carried out at the
applied voltage $U$ = 60~V for SiO${ }_2$(Co) films and at $U$ = 3~V
for the SiO${ }_2$(Co)/Si structure. The magnetic field $H$ = 10~kOe
is parallel to the surface of the granular film. It can be seen that
temperature decreasing causes to the growth of the absolute value of
the intrinsic magnetoresistance for SiO${ }_2$(Co) films. For the
SiO${ }_2$(Co)/Si structure electrons are injected from the granular
film into the semiconductor and temperature decreasing leads to the
change of the magnetoresistance sign.

\begin{figure*}
\begin{center}
\includegraphics*[scale=.45]{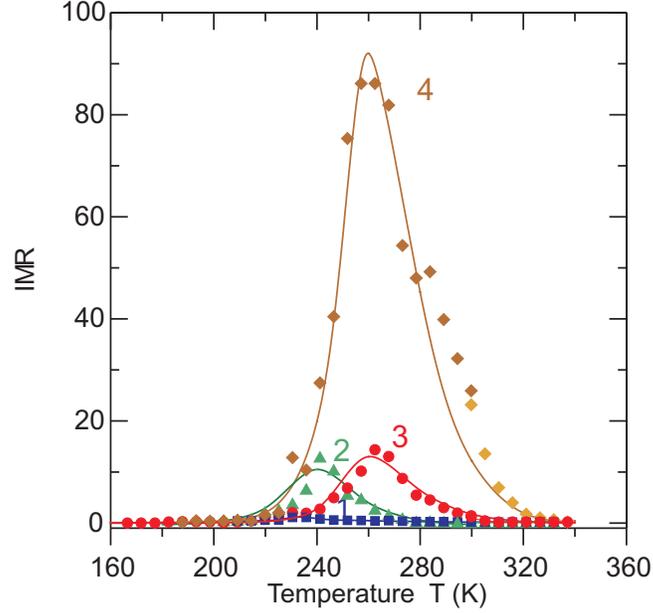}
\end{center}
\caption{Temperature dependencies of the injection
magnetoresistance, IMR, for the SiO${ }_2$(Co)/GaAs structure with
$x$ = 71~at.\% Co content in the in-plane magnetic field $H$ =
10~kOe at applied voltages: (1) $U$ = 40~V, (2) 50~V, (3) 60~V, (4)
70~V. Solid lines are theoretical fittings. } \label{Fig8}
\end{figure*}

\begin{figure*}
\begin{center}
\includegraphics*[scale=.45]{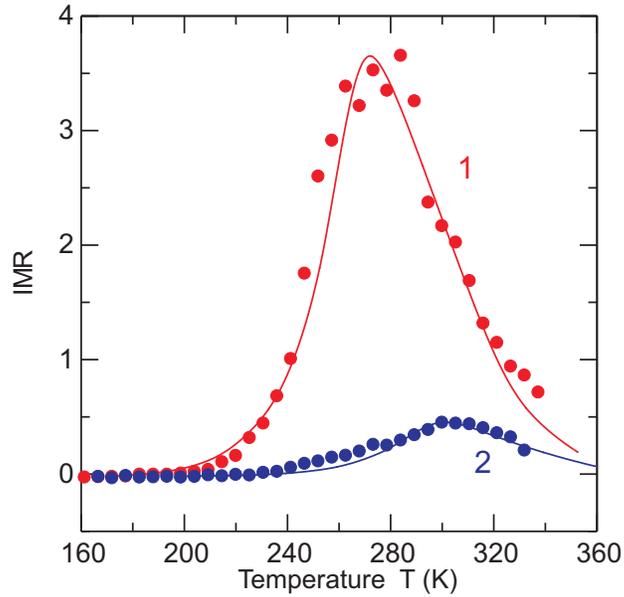}
\end{center}
\caption{Temperature dependencies of the magnetoresistance, IMR, for
the SiO${ }_2$(Co)/GaAs structure with $x$ = 38~at.\% Co content  in
the in-plane magnetic field $H$ = 10~kOe at the applied voltage $U$
= 60~V. (1) Electrons are injected from the SiO${ }_2$(Co) film into
GaAs, (2) electrons drift from GaAs into the granular film. Solid
lines are theoretical fittings.} \label{Fig9}
\end{figure*}

Temperature dependencies of the IMR for SiO${ }_2$(Co)/GaAs
structures essentially differ from the above-mentioned dependencies
for SiO${ }_2$(Co)/GaAs structures and SiO${ }_2$(Co) films. They
have a peak type character (Figures \ref{Fig8} and \ref{Fig9}). The
temperature location of the peak depends on the Co concentration and
can be shifted by the applied electrical field. Figure \ref{Fig8}
shows temperature dependencies of the IMR for SiO${ }_2$(Co)/GaAs
with 71~at.\% Co at different applied voltages, when electrons are
injected from the granular film into the GaAs substrate. Increasing
voltage $U$ causes to a shift of the peak to higher temperatures. At
the same time, the voltage growth leads to an increase of the peak
magnitude. For SiO${ }_2$(Co)/GaAs structure with lower Co content
($x$ = 38~at.\%, Figure \ref{Fig9}), the temperature peak of the IMR
has higher value of width. For the case, when electrons move from
GaAs into the SiO${ }_2$(Co) film, the IMR peak is located at higher
temperature and its magnitude is lower.

\section{Theoretical model and explanation of experimental results}
\subsection{Theoretical model}

Explanation of the IMR effect is based on the theoretical model of
the magnetic-field-controlled avalanche process triggered by
electrons passed through the spin-dependent potential barrier in the
accumulation layer in the SC at the interface. The applied magnetic
field reduces the transparency of the spin-dependent potential
barrier. This leads to a decrease of the kinetic energy of injected
electrons and to the suppression of the impact ionization onset.

\begin{figure*}
\begin{center}
\includegraphics*[scale=.5]{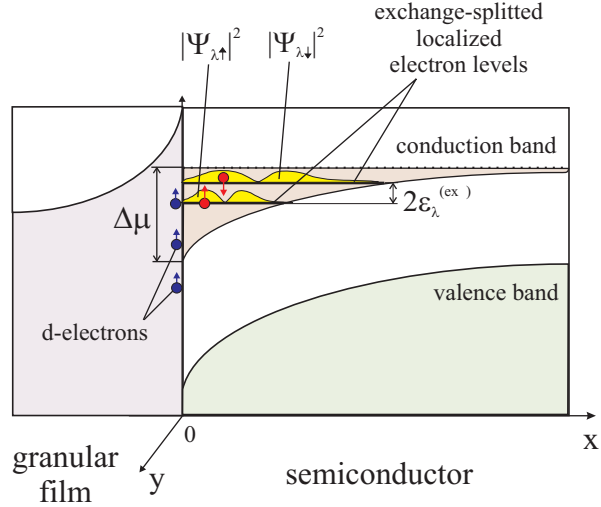}
\end{center}
\caption{Electronic energy band structure at the contact region of
the ferromagnet / semiconductor.} \label{Fig10}
\end{figure*}

Let us consider formation of the accumulation electron layer in the
SC at the interface, the spin-dependent potential barrier and the
IMR effect caused by the barrier~\cite{ref40}. In the FM/SC
heterostructure the difference of chemical potentials $\Delta\mu$
between the FM and the SC determines bending of the SC conduction
band (Figure \ref{Fig10}). $d$-electrons in the FM at the interface
and electrons in the accumulation electron layer in the SC are
coupled by the exchange interaction $J_0(\vec{r}-\vec{R})$. The
Hamiltonian of the model is written in the form

\[{\bfb H} = {\bfb H}_e +{\bfb H}_{ed} +{\bfb H}_{\varphi},\]

\noindent where
\[{\bfb H}_e = \sum_{\alpha}\int \Psi^{+}_{\alpha}(\vec{r})
\left[-\frac{\hbar^2}{2m}\Delta -\mu - e\varphi
(\vec{r})\right]\Psi_{\alpha}(\vec{r})\,d\vec{r}\]

\noindent is the Hamiltonian of electrons with the mass $m$ and the
charge $e$ in the SC in the electrical field with the potential
$\varphi (\vec{r})$. $\mu$ is the chemical potential.
$\Psi^{+}_{\alpha}(\vec{r})=\sum_{\lambda}\psi^{*}_{\lambda}(\vec{r})
a^{+}_{\lambda\alpha}$, $\Psi_{\alpha}(\vec{r})=
\sum_{\lambda}\psi_{\lambda}(\vec{r}) a_{\lambda\alpha}$ are the
second-quantized wavefunctions of an electron with a spin $\alpha$ =
$\uparrow ,\downarrow$. $a^{+}_{\lambda\alpha}$, $a_{\lambda\alpha}$
are the creation and annihilation Fermi operators, respectively, for
an electron with the wavefunction $\psi_{\lambda}(\vec{r})$ with the
multiindex $\lambda$.

\[{\bfb H}_{ed} = -\sum_{\vec{R}}\int J_0(\vec{r}-\vec{R})(\vec{S}(\vec{R}),
\vec{\sigma}(\vec{r}))\,d\vec{r}\]

\noindent is the exchange interaction Hamiltonian between the spin
density $\vec{\sigma}(\vec{r})$ of electrons in the SC and spins
$\vec{S}(\vec{R})$ of $d$-electrons in the FM. The vector spin
density operator $\vec{\sigma}(\vec{r})$ is determined by operators
$\Psi_{\alpha}(\vec{r})$, $\Psi^{+}_{\alpha}(\vec{r})$

\[\sigma_x(\vec{r}) = \Psi^{+}_{\uparrow}(\vec{r})
\Psi_{\downarrow}(\vec{r}) + \Psi^{+}_{\downarrow}(\vec{r})
\Psi_{\uparrow}(\vec{r})\]

\[\sigma_y(\vec{r}) = -i\Psi^{+}_{\uparrow}(\vec{r})
\Psi_{\downarrow}(\vec{r}) + i\Psi^{+}_{\downarrow}(\vec{r}) \Psi_{\uparrow}(\vec{r}) \]

\[\sigma_z(\vec{r}) = \Psi^{+}_{\uparrow}(\vec{r})
\Psi_{\uparrow}(\vec{r}) - \Psi^{+}_{\downarrow}(\vec{r}) \Psi_{\downarrow}(\vec{r}) .\]

\noindent The Hamiltonian
\[{\bfb H}_{\varphi} = -\frac1{8\pi}\int [\nabla\varphi(\vec{r})]^2 \,d\vec{r}\]

\noindent describes the classical inner electrostatic field
$\varphi(\vec{r})$.

In order to find the effective exchange interaction between spins
$\vec{S}(\vec{R})$ of $d$ electrons in the FM and the spin
$\vec{\sigma}^{(in)}(\vec{r})$ of an injected electron with the
wavefunction $\psi^{(in)}_{\alpha}(\vec{r})$ ($\alpha$ = $\uparrow
,\downarrow$) and the spin-dependent potential barrier, the
temperature diagram technique is used~\cite{ref55,ref56}. Before
this we consider formation of the accumulation electron layer.

\subsubsection{Formation of the accumulation electron layer}

In the self-consistent-field approximation of the diagram expansion
electrons of the conduction band in the SC and the inner
self-consistent electrical field are described by the following
equations.

\noindent (1) Equation for the electron wavefunction in the SC

\begin{equation}
\left[-\frac{\hbar^2}{2m}\frac{d^2}{dx^2} - e\varphi (x)\right]
\chi_{\nu}(x)= \varepsilon_{\nu}^{(0)} \chi_{\nu}(x), \label{eq2}
\end{equation}

\noindent where $\psi_{\lambda}(\vec{r})= V^{-1/2}\chi_{\nu}(x)
\exp(iq_yy +iq_zz)$ is the electron wavefunction in the volume $V$
of the SC with the multiindex $\lambda =(\nu,q_y,q_z)$ and the
energy spectrum $\varepsilon_{\lambda}= \varepsilon_{\nu}^{(0)}+
\hbar^2(q^2_y + q^2_z)/2m$.

\begin{figure*}
\begin{center}
\includegraphics*[scale=.5]{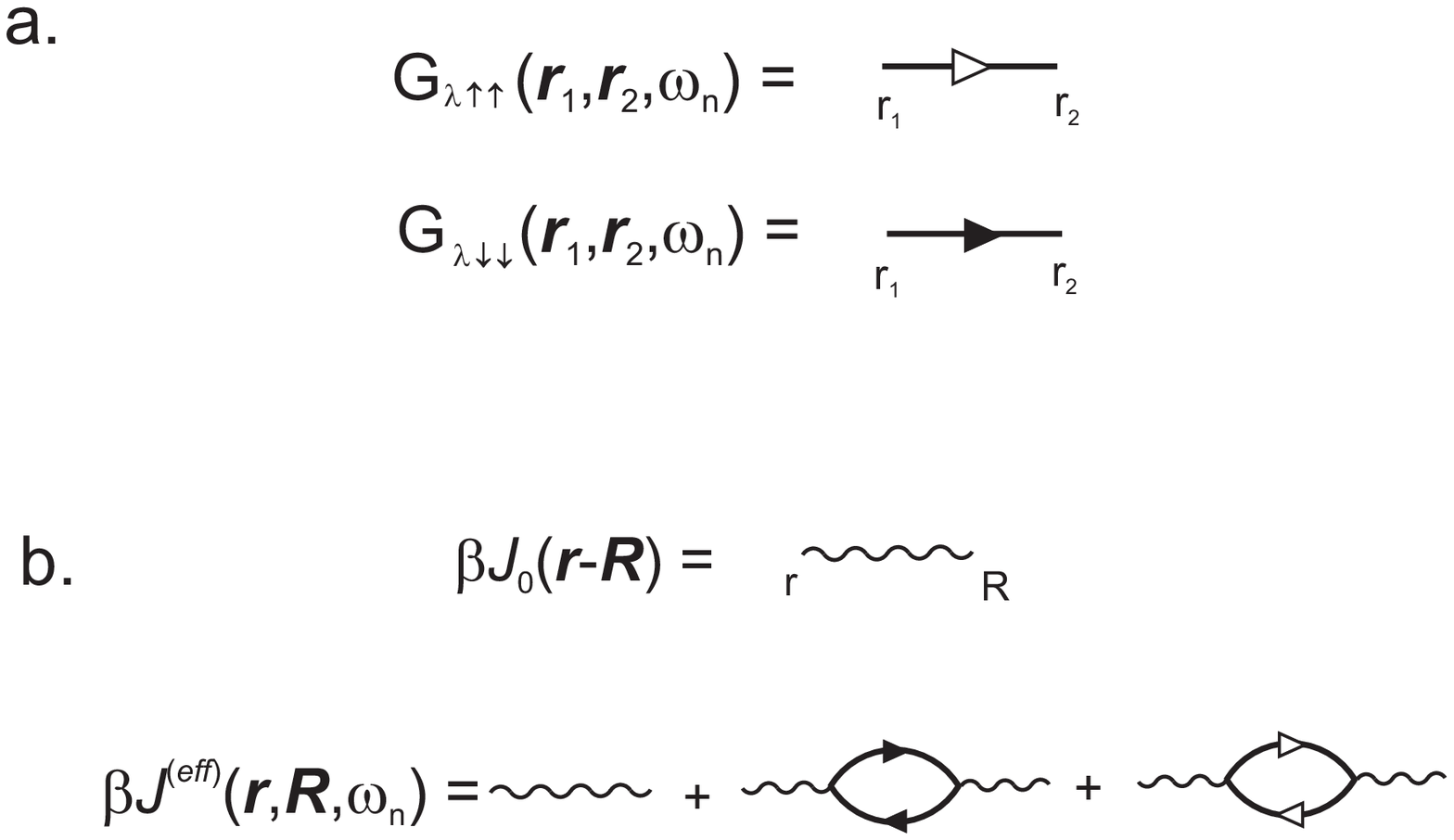}
\end{center}
\caption{(a) Temperature electron Green functions with the spin
$\uparrow$ and $\downarrow$. (b) Bare and effective exchange
interactions. } \label{Fig11}
\end{figure*}

\noindent (2) The equation for the inner self-consistent electrical
field

\[ \Delta\varphi(\vec{r})= 4\pi e\left\{\sum_{\lambda,\omega_n}\left[
G_{\lambda \uparrow \uparrow}( \vec{r}, \vec{r},\omega_n) +
G_{\lambda \downarrow \downarrow}( \vec{r},
\vec{r},\omega_n)\right.\right.\]
\begin{equation}
- \left.\left.G^{(0)}_{\lambda \uparrow \uparrow}( \vec{r},
\vec{r},\omega_n) - G^{(0)}_{\lambda \downarrow \downarrow}(
\vec{r}, \vec{r},\omega_n)\right]\right\} , \label{eq3}
\end{equation}

\noindent where

\begin{equation}
G_{\lambda{\alpha}_1{\alpha}_2}( \vec{r}_1, \vec{r}_2,\omega_n)=
\frac{\psi^{*}_{\lambda}(\vec{r}_1)
\psi_{\lambda}(\vec{r}_2)\delta_{{\alpha}_1{\alpha}_2}}{\beta
(i\hbar\omega_n  - E_{\lambda\alpha_1}+\mu)}, \label{eq4}
\end{equation}

\noindent are electron Green functions (Figure~\ref{Fig11}(a)),
$\beta =1/kT$, $k$ is the Boltzmann constant, $T$ is the
temperature, $\hbar\omega_n = (2n+1)\pi/\beta$,  $n$ is an integer,

\begin{equation}
E_{\lambda\alpha}= \varepsilon_{\lambda} \mp
\varepsilon_{\lambda}^{{\rm (ex)}}. \label{eq5}
\end{equation}

\noindent The upper sign in equation~(\ref{eq5}) corresponds to
$\alpha$ = $\uparrow$; the lower sign, to $\alpha$ = $\downarrow$.
The energy $\varepsilon_{\lambda}^{{\rm (ex)}}$ is determined by the
exchange Hamiltonian ${\bfb H}_{ed}$ in the self-consistent-field
approximation

\begin{equation}
\varepsilon_{\lambda}^{{\rm (ex)}} = -\sum_{\vec{R}}\int
J_0(\vec{r}-\vec{R})(\langle\vec{S}(\vec{R})\rangle_0,
\langle\vec{\sigma}(\vec{r})\rangle_0) \,d\vec{r}. \label{eq6}
\end{equation}

\noindent $\langle\vec{S}(\vec{R})\rangle_0$ and
$\langle\vec{\sigma}(\vec{r})\rangle_0$ are the statistical-average
$d$-electron spin in the FM and the electron spin density in the SC,
respectively. $G^{(0)}_{\lambda \alpha \alpha}$ are electron Green
functions determined in the single SC in the absence of the
electrical field.

\noindent (3) The relationship between the chemical potential $\mu$
and the electron concentration $n_0$ in the single SC

\begin{equation}
n_0 = \frac{8\pi e}{V}\sum_{\vec
q=(q_x,q_y,q_z)}n_F[\beta(\hbar^2|\vec q|^2/2m - \mu)] .\label{eq7}
\end{equation}

\noindent where $n_F(a)= [\exp(a)+1]^{-1}$.

Equations (\ref{eq2}), (\ref{eq3}), (\ref{eq7}) are simultaneous
equations in unknowns: the wave function $\chi_{\nu}(x)$, the energy
$\varepsilon^{(0)}_{\nu}$, the electrical potential $\varphi (x)$,
and the chemical potential $\mu$ in the SC. Taking into account that
at the interface of the heterostructure ($x = 0$) the potential
$\varphi(x)$ is determined by the difference of chemical potentials
$\Delta\mu$ between the SC and the FM, $\varphi(0) = \Delta\mu /e$,
and at a great distance from the interface, when $x \to\infty$, the
potential $\varphi(x)$ tends to zero, we numerically can solve
equations (\ref{eq2}), (\ref{eq3}), (\ref{eq7}).

\subsubsection{The effective exchange interaction and the spin-dependent potential
barrier}

The effective exchange interaction and the spin-dependent potential
barrier for injected electrons are found in the next approximation
of the diagram expansion. This is the one-loop approximation with
respect to the bare exchange interaction $J_0(\vec r -\vec R)$
(figure~\ref{Fig11}(b)). In this approximation we take into account
solutions of equations (\ref{eq2}), (\ref{eq3}), (\ref{eq7}) made in
the self-consistent-field approximation and find the effective
exchange interaction

$$J^{\rm (eff)}(\vec r,\vec R,\omega_n) = J_0(\vec{r}-\vec{R}) +
J_1(\vec r,\vec R,\omega_n) ,$$

\noindent where the interaction $J_1$ has the form

\[ J_1(\vec r,\vec R,\omega_n) = -\beta\int\!\int J_0(\vec{r}-\vec{r}_1)
\sum_{k,\lambda_1,\lambda_2}\left[G_{\lambda_1 \uparrow \uparrow}(
\vec{r}_1, \vec{r}_2, \omega_k) G_{\lambda_2 \uparrow \uparrow}(
\vec{r}_1, \vec{r}_2, \omega_k + \omega_n)\right.\]
\begin{equation}+ \left. G_{\lambda_1 \downarrow \downarrow}(
\vec{r}_1, \vec{r}_2, \omega_k) G_{\lambda_2 \downarrow \downarrow}(
\vec{r}_1, \vec{r}_2, \omega_k + \omega_n)\right]
J_0(\vec{r}_2-\vec{R}) \, d\vec{r}_1 d\vec{r}_2.\label{eq8}
\end{equation}

\begin{figure*}
\begin{center}
\includegraphics*[scale=.4]{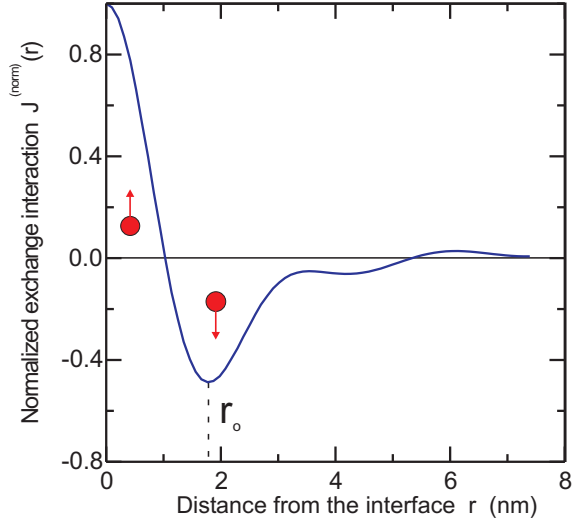}
\end{center}
\caption{Normalized exchange interaction $J^{\rm (norm)}(r)=
J_1(\vec r,0,0) / J_1(0,0,0)$ for the heterostructure with the
difference of chemical potentials $\Delta\mu$ = 150~meV and with the
electron concentration $n_0$  = 1$\times$10${ }^{15}$~cm${ }^{-3}$
in the SC at $T$ = 300~K. } \label{Fig12}
\end{figure*}

In the relation (\ref{eq8}) the Green functions
$G_{\lambda{\alpha}_1{\alpha}_2}$ (\ref{eq4}) are expressed via
wavefunctions $\psi_{\lambda}(\vec{r})$, the chemical potential
$\mu$ and the electron energy $E_{\lambda\alpha}$ (\ref{eq5}). The
interaction $J_1$ is of the RKKY-type (Ruderman, Kittel, Kasuya,
Yosida~\cite{ref57,ref58,ref59}). Spins of electrons in the
accumulation layer shield spins of $d$-electrons in the FM at the
interface. As the result of this shielding, the short-range exchange
interaction $J_0(\vec{r}-\vec{R})$ is transformed into the
long-range effective exchange interaction $J^{\rm (eff)}(\vec r,\vec
R,\omega_n)$, which changes its sign at a some distance from the
interface (figure~\ref{Fig12}). To find the numerical solution, we
assume that $ J_0(\vec{r} -\vec{R}) = J_0\exp(-\xi|\vec{r}-
\vec{R}|)$ in equations (\ref{eq6}), (\ref{eq8}), where $\xi$ is the
reciprocal radius of the exchange interaction and $J_0$ is
determined by the Coulomb interaction with $d$-electrons on a FM
atom~\cite{ref60}. Calculations have been drawn, when $\omega_n =0$,
$\vec R=0$, $\xi$ = 10 nm${ }^{-1}$, $J_0$  = 2~eV,
$|\langle\vec{S}( \vec{R})\rangle_0| = 1/2$,
$|\langle\vec{\sigma}(\vec{r})\rangle_0| =
1/2|\psi_{\lambda}(\vec{r})|^2$, $\Delta\mu$ = 150~meV, $n_0$  =
1$\times$10${ }^{15}$~cm${ }^{-3}$ at $T$ = 300~K for the cubical
crystal FM lattice with the lattice constant $a$ = 0.23 nm. At the
distance $r_0$ the exchange interaction $J_1$ has a maximum opposite
value. If the accumulation layer (quantum well) contains a great
number of electron states, the distance $r_0$ can be evaluated as
the half of the period of the Ruderman-Kittel function, $r_0\approx
\frac12(\pi/3n_s)^{1/3} $~\cite{ref57,ref58,ref59}, where $n_s$ is
the electron density at the interface.

In order to find the spin-dependent potential barrier, we assume
that the magnetic field $\vec H$ is parallel to the axis Oz. Then,
the height of the energy barrier formed by the effective exchange
interaction for injected spin-polarized electrons, which move from
the interface, is determined by the relation

\begin{equation}
W = \sum_{\vec{R}}\int \langle\sigma^{(in)}_z(\vec{r})\rangle J^{\rm
(eff)}(\vec r,\vec R,0) \langle S_z(\vec{R})\rangle_0 \,d\vec{r},
\label{eq9}
\end{equation}

\noindent where $\langle\sigma^{(in)}_z( \vec{r})\rangle =\langle
\psi^{(in)*}_{\uparrow}(\vec{r}) \psi^{(in)}_{\uparrow}(\vec{r})
-\psi^{(in)*}_{\downarrow}(\vec{r}) \psi^{(in)}_{\downarrow}
(\vec{r}) \rangle $, $\langle S_z(\vec{R})\rangle_0$ is the
$z$-projection of the statistical-average $d$-electron spin at the
site $\vec R$ at the interface. For calculation of $W$ we assume
that the spin density $\langle\sigma^{(in)}_z( \vec{r})\rangle
=1/2\cdot\delta (r-r_0)$. We have found, that, if the accumulation
layer contains a small number of localized electron states
$\chi_{\nu}(x)$, which are determined by equation (\ref{eq2}), then
these states give the main contribution to the exchange interaction
$J_1$ in equation~(\ref{eq8}) and to the height of the energy
barrier $W$~(\ref{eq9}). The maximum of the barrier is observed,
when the accumulation layer has two sublevels of an
exchange-splitted localized electron state (Figure~\ref{Fig10}).
Exchange-splitted localized states have high values of the exchange
energy $\varepsilon_{\lambda}^{{\rm (ex)}}$~(\ref{eq6}) and this
causes to high values of the barrier $W$~(\ref{eq9}). If the
accumulation layer does not contain localized states, the magnitude
of $W$ sharply falls. Dependencies of $W$ on the difference of
chemical potentials $\Delta\mu$ and temperature dependencies are
presented in~\cite{ref40}.

\subsubsection{The IMR effect}

\begin{figure*}
\begin{center}
\includegraphics*[scale=.6]{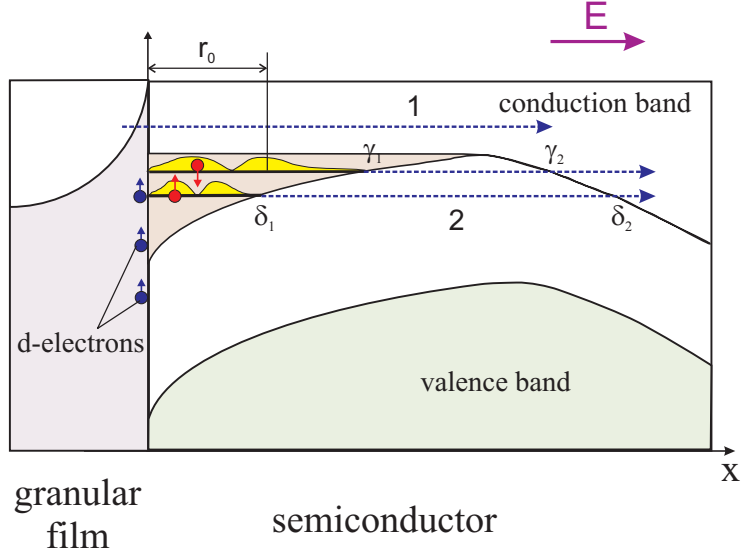}
\end{center}
\caption{Schematic band diagram at the applied electrical field at
the contact region of the ferromagnet / semiconductor. (1)
Surmounting injected electrons over the spin-dependent potential
barrier formed by localized states at the distance $r_0$ from the
interface, (2) tunneling from exchange-splitted localized states. }
\label{Fig13}
\end{figure*}

\begin{figure*}
\begin{center}
\includegraphics*[scale=.6]{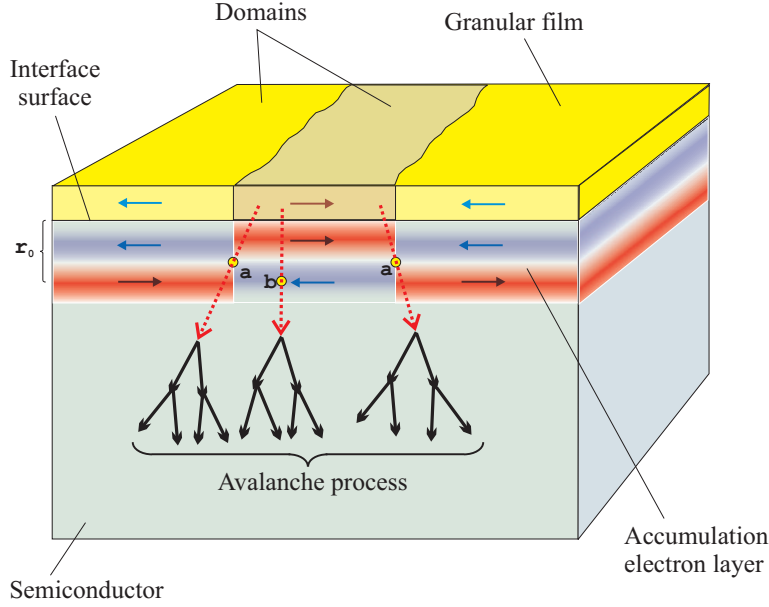}
\end{center}
\caption{Trajectories without spin-flip scattering of injected
electrons and without surmounting the potential barrier on the
accumulation layer (points $a$) and trajectories with surmounting
the potential barrier (points $b$). } \label{Fig14}
\end{figure*}

The observed IMR effect can be explained by the developed
theoretical model. Applied electrical field bends the SC conduction
band (Figure~\ref{Fig13}). Two ways of the spin-polarized current
injected into the SC can be supposed: (1) injected electrons
surmount the spin-dependent potential barrier $W$ at the distance
$r_0$ from the interface, (2) spin-polarized electrons tunnel from
sublevels of the exchange-splitted localized states. Let us consider
the first way. In the absence of an external magnetic field, the
domain structure of the granular film (Figure~\ref{Fig1}) induces
corresponding spin orientations of electrons localized in the
accumulation layer and this domain structure has domain walls
(Figure~\ref{Fig14}). In this case, electrons injected from the
granular film can cross through the accumulation layer without a
loss of their spin polarization and without surmounting the
potential barrier on channels close to domain walls (trajectories
with points $a$). In the magnetic field of high values, when domains
disappear, spin polarized electrons moving in the SC from the
interface must surmount the potential barrier at the distance $r_0$
(trajectories with points $b$). Electrons surmounted the potential
barrier trigger the process of impact ionization. According
to~\cite{ref61}, the voltage drop is concentrated mainly in the
vicinity of the barrier. Consequently, the avalanche process
originates in this vicinity region. The current density $j$ flowing
in the heterostrucure is determined by the concentration of injected
electrons $n$, the average velocity $v$ and the multiplication
factor $M$ of the avalanche process, $j$ = $Menv$. We suppose that
in the absence of a magnetic field electrons cross through the
points $a$ and in the magnetic field electrons surmount the barrier
through the points $b$. Then, taking into account that
$n_a=n_{int}\exp(eU_a/ kT)$, $n_b=n_{int}\exp[(eU_b-W)/kT]$, where
$n_{int}$ is the electron concentration at the interface at the
Fermi level, $U_a$ and $U_b$ are differences of potentials between
the interface and points $a$ and $b$, respectively, from equation
(\ref{eq1}) for the tunnel opaque potential barrier we get

\begin{equation}
\mbox{\emph{IMR}}(W,T)= \frac{M_an_av_a}{M_bn_bv_b}-1
=A\exp\left(\frac{W}{kT}\right)-1 .\label{eq10}
\end{equation}

\noindent The coefficient $A$ is equal to $M_av_a/M_bv_b\exp[e(U_a-
U_b)/kT]$. It is need to notice that the relation (\ref{eq10}) is
truthful for high values of the magnetic field, when domains
disappear. Taking into account that the energy barrier $W$ sharply
depends on temperature~\cite{ref40}, in the first approximation the
temperature dependence of the IMR is determined by the term
$\exp(W/kT)$.

In the second case of the injected spin-polarized current, electrons
with the spin $\alpha$ and the energy $E_{\lambda\alpha}$
(\ref{eq5}) tunnel from sublevels of the exchange-splitted localized
states. In the absence of an external magnetic field because of the
existence of the domain structure in the granular film, spin
orientations of localized electrons on sublevels in the accumulation
layer are different for neighbouring regions corresponding to
domains. In this case, tunneling occurs on channels close to domain
walls. In the magnetic field of high values, domains disappear and
the tunneling transparency decreases. The coefficient of the
tunneling transparency from the state $E_{\lambda\alpha}$
is~\cite{ref62}

$$D_{\lambda\alpha}= \exp{\left\{-\frac2{\hbar}
\int^{\gamma_2(\delta_2)}_{\gamma_1(\delta_1)}
\left|[2m(\bar{V}(x)-E_{\lambda\alpha})]^{1/2}
\right|\,dx\right\}},$$

\noindent where $\bar{V}(x)$ is the potential energy of the barrier.

It is need to notice that the magnetic field can act on the
avalanche process directly. The application of a magnetic field
increases the density of states at the bottom of the lowest Landau
level, causing electrons to occupy states with lower
energy~\cite{ref37}. Electrons orbitals become more localized in the
vicinity of donor ions and the overlap by their tails is reduced.
Accordingly, the current decreases since fewer electrons can take
part in the impact ionization process. The second mechanism of the
magnetic-field action is quasi-neutrality breaking of the
space-charge effect, where insufficient charge is present to
compensate electrons injected into the SC~\cite{ref63}. The
above-mentioned factors do not explain the observed temperature-peak
type dependence of the IMR effect, but they can enhance the IMR
value.

\subsection{Explanation of the experiment}

In order to explain high values of the IMR effect in SiO${
}_2$(Co)/GaAs heterostructures and the temperature-peak type
character, we use the developed theoretical model. The developed
theory can be applied for these heterostructures, if the size of Co
nanoparticles is less than the thickness $l$ of the accumulation
layer. In this case, the granular film can be considered as
continuous and can be characterized by statistical-average
parameters. The thickness $l$ depends on the difference of the
chemical potentials $\Delta\mu = \mu_g - \mu_s $, where $\mu_s$ is
the chemical potential in the SC and $\mu_g$ is the chemical
potential in the granular film. In the first approximation, the
chemical potential $\mu_g$ is given by

\begin{equation}
\mu_g = \mu_{{\rm SiO}_2}(x_{{\rm SiO}_2}/100) + \mu_{\rm Co}(x_{\rm
Co}/100) ,\label{eq11}
\end{equation}

\noindent where $\mu_{{\rm SiO}_2}$, $\mu_{\rm Co}$ are the chemical
potentials of the SiO${ }_2$ matrix and Co nanoparticles; $x_{{\rm
SiO}_2}$, $x_{\rm Co}$ are the atomic concentrations of the SiO${
}_2$ and Co in percents, respectively. The difference $\Delta\mu$
between chemical potentials of the GaAs and the SiO${ }_2$(Co)
granular film and between chemical potentials of the Si substrate
and the granular film can be estimated from well known values of the
energy of the thermoelectron emission. For the given materials the
differences of the chemical potentials are $\mu_{{\rm SiO}_2}-
\mu_{\rm{Co}}$ = 0.59~eV, $\mu_{{\rm SiO}_2}- \mu_{\rm{GaAs}}$ =
0.62~eV, $\mu_{\rm Co}- \mu_{\rm GaAs}$ = 0.03~eV, $\mu_{{\rm
SiO}_2}- \mu_{\rm{Si}}$ = 0.95~eV, $\mu_{\rm Co}- \mu_{\rm Si}$ =
0.36~eV~\cite{ref64}.

In order to solve equations (\ref{eq2}), (\ref{eq3}), (\ref{eq7}) in
the approximation of the continuous granular film model, we need to
find the surface probability of the Co particle distribution at the
interface. We assume that at the interface Co particles are randomly
allocated with the surface probability

$$s=p^{2/3}=\left[\frac{x_{\rm Co}v_{\rm Co}}{x_{\rm Co}v_{\rm Co}
+(100-x_{\rm Co})v_{{\rm SiO}_2}}\right]^{2/3} ,$$

\noindent where $p$ is the relative Co volume, $v_{\rm Co}=m_{\rm
Co}/\varrho_{\rm Co}N_A$, $v_{{\rm SiO}_2}=m_{{\rm SiO}_2}/
\varrho_{{\rm SiO}_2}N_A$ are atomic and molecular volumes for the
Co and the SiO${ }_2$ matrix; $m_{\rm Co}$, $m_{{\rm SiO}_2}$ are
the respective atomic and molecular masses; $\varrho_{\rm Co}$,
$\varrho_{{\rm SiO}_2}$ are the densities of Co particles and the
SiO${ }_2$ matrix; $N_A$ is the Avogadro number. For calculations we
use $m_{\rm Co}$ = 58.93~a.m., $m_{{\rm SiO}_2}$ = 60.09~a.m.,
$\varrho_{\rm Co}$ = 8.90~g/cm${ }^3$, $\varrho_{{\rm SiO}_2}$ =
2.26~g/cm${ }^3$~\cite{ref65}. According to the continuous granular
film approximation, we must made substitutions $\langle\vec{S}(
\vec{R})\rangle_0\to s\langle\vec{S}(\vec{R})\rangle_0$ and
$\langle{S_z}(\vec{R})\rangle_0\to s\langle{S_z}(\vec{R})\rangle_0$
in relations (\ref{eq6}) and (\ref{eq9}), respectively.

Using the developed model, we have found the electron wavefunction
$\chi_{\nu}(x)$ (\ref{eq2}), the inner self-consistent electrical
field $\varphi(\vec{r})$ (\ref{eq3}), and the energy barrier $W$
(\ref{eq9}). Calculations have been made for the effective exchange
interaction $J_0(\vec{r} -\vec{R}) = J_0\exp(-\xi|\vec{r}-\vec{R}|)$
with $J_0$ = 2~eV, $\xi$ = 2~nm${ }^{-1}$~\cite{ref60}. For SiO${
}_2$(Co)/GaAs heterostructures at the given temperatures 160 -
340$^{\circ}$C the thickness $l$ of the accumulation layer is in the
range 8 - 50~nm. The size of Co nanoparticles is less than the
thickness $l$, and the approximation of the continuous granular film
is truthful. Heterostructures possess localized electron states in
the accumulation layer at the interface. In contrast, for SiO${
}_2$(Co)/Si heterostructures due to higher values of the difference
of the chemical potentials $\Delta\mu$ at the interface the
potential depth of the accumulation layer is deeper. This leads to
higher electron concentration at the interface and to more efficient
shielding of Co spins. As a result of this, the accumulation layer
has small thickness without any localized states. The absence of
localized states in SiO${ }_2$(Co)/Si heterostructures explains
small values of the barrier $W$ (\ref{eq9}) and small values of the
IMR effect (figures~\ref{Fig6},~\ref{Fig7}) in comparison with IMR
values in SiO${ }_2$(Co)/GaAs heterostructures (figures~\ref{Fig4},
\ref{Fig5}, \ref{Fig8}, \ref{Fig9}). Let us consider IMR
dependencies on the Co concentration, temperature and the magnetic
field.

\subsubsection{IMR dependence on the Co concentration}

The dependence of the IMR on the Co concentration $x$ for SiO${
}_2$(Co)/GaAs structures, when electrons are injected from the SiO${
}_2$(Co) film (figure~\ref{Fig5}), demonstrates high IMR values for
the concentration range $x$ = 54 - 71~at.\% and low IMR values for
lower and higher Co concentrations. From the developed model it is
found that structures with $x$ = 54 - 71~at.\% have one - two
electron localized states with high energies $\varepsilon_{\lambda
}^{{\rm (ex)}}$~(\ref{eq6}), which leads to high barrier $W$ at room
temperature. Heterostructures with lower Co concentration possess
greater number of localized states in the accumulation layer with
energies $\varepsilon_{\lambda }^{{\rm (ex)}}$ of small values. For
these structures the IMR coefficient is low. If the Co concentration
$x >$ 71~at.\%, the accumulation layer has small thickness without
localized states and is transparent for current.

\subsubsection{Temperature dependencies of the IMR}

At the interface the electron concentration increases with
temperature increasing. At low temperatures the accumulation layer
contains large number of exchange-splitted localized states with
small energies $\varepsilon_{\lambda }^{{\rm (ex)}}$. Temperature
increasing induces thinning of the accumulation layer, a decrease of
the localized state number, an increase of energies
$\varepsilon_{\lambda}^{{\rm (ex)}}$, and a growth of the barrier
$W$. At a certain temperature the accumulation layer contains one
exchange-splitted level and the magnitude of $W$ reaches the maximum
value. The further temperature growth gives higher electron
concentration at the interface, more efficient shielding of Co
spins, and thinner thickness of the accumulation layer. When the
sublevel, on which electrons have spin orientations opposite to Co
spins, crosses the Fermi level, the height of the potential barrier
$W$ sharply decreases. In Figure~\ref{Fig2} for the SiO${
}_2$(Co)/GaAs structure with the Co content 71~at.\% crossing of the
Fermi level is manifested as a fall on the temperature dependence of
the inject current at $T$ = 320~K at the applied voltage $U$ = 70~V.
This fall in the current corresponds to the disappearance of the IMR
effect at 320~K, $U$ = 70~V in Figure~\ref{Fig8}.

The temperature-peak type character of the IMR effect is presented
in figures~\ref{Fig8},~\ref{Fig9}. Maxima of peaks correspond to one
exchange-splitted level in the accumulation layer. Neglecting
spin-polarized tunneling from exchange-splitted localized states, we
fit experimental results using the relation~(\ref{eq10}). The
barrier $W$ is given by equation (\ref{eq9}) and the amplitude $A$
in the relation~(\ref{eq10}) is determined to reach the best fit of
the peak height. According to the developed model, the peak width is
inversely proportional to the magnitude of the surface probability
$s$ of the Co particle distribution at the interface. Decreasing the
Co content results in the decrease of the surface probability $s$:
from $s$ = 0.52 ($x$ = 71~at.\% Co) to $s$ = 0.26 ($x$ = 38~at.\%
Co). This corresponds to the observed increase of the peak width
with Co concentration decreasing: from $\Delta T$ = 37~K ($x$ =
71~at.\% Co) to $\Delta T$ = 62~K ($x$ = 38~at.\% Co).

Locations of IMR temperature peaks can be shifted by the applied
electrical field. These shifts can be explained by the change of the
electron concentration at the interface under the electrical field
action. The applied field causes to an electron depletion in the SC
at the interface. As a result of this, at high field magnitudes it
is need higher temperatures to form the accumulation layer with one
exchange-splitted level. In order to take into account the action of
the electrical field for the SiO${ }_2$(Co)/GaAs structure with the
Co content 71~at.\% in Figure \ref{Fig8}, we use the following
differences of the chemical potentials $\Delta\mu$: 0.201~eV ($U$ =
40~V), 0.197~eV ($U$ = 50~V), 0.187~eV ($U$ = 60~V and 70~V).

\subsubsection{IMR dependencies on the magnetic field}

\begin{figure*}
\begin{center}
\includegraphics*[scale=.6]{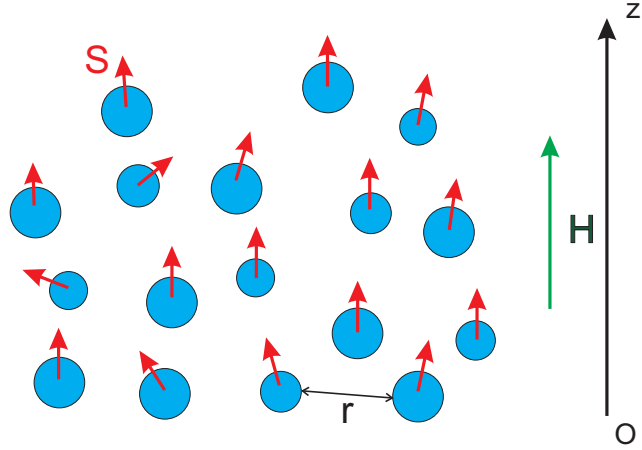}
\end{center}
\caption{Alignment of spins of Co particles at the interface along
the magnetic field direction. Distance $r$ between Co particles is a
random parameter. } \label{Fig15}
\end{figure*}

At last, we consider IMR dependencies on the magnetic field for
SiO${ }_2$(Co)/GaAs structures. As we can see from Figure
\ref{Fig4}, at magnetic fields of low values the IMR grows greater
than at high magnetic fields. The high growth of the IMR can be
explained by changes of the domain structure, which disappears at
$H\approx$ 3 - 4~kOe. Slow IMR increasing at magnetic fields of high
values can be due to alignment of different spin orientations of
randomly allocated Co particles at the interface (Figure
\ref{Fig15}). Magnetic field polarizes spins along the field
direction. This leads to the increase of the $z$-projection
$\langle{S_z}(\vec{R})\rangle_0$ and to the increase of the barrier
height $W$ (\ref{eq9}).

\section{FM / SC heterostructures with quantum wells with
spin-polarized localized electrons}

Considering the IMR effect in SiO${ }_2$(Co)/GaAs heterostructures,
we can result in conclusion that for the efficient magnetoresistance
and spin injection in FM / SC heterostructures it is need to fulfill
the following requirements.

\noindent (1) The SC contains a quantum well at the interface.

\noindent (2) The quantum well must contain localized electron
levels.

\noindent (3) Localized levels must be exchange-splitted by the FM.

\noindent (4) Giant magnetoresistance effect can be achieved on the
base of the avalanche breakdown phenomena.

It is very important to extend the IMR effect observed in SiO${
}_2$(Co)/GaAs heterostructures to heterostructures with other SCs
and to reach high efficient spin-polarized electron injection. One
of the promising semiconductor for spintronics with enhanced
lifetime and transport length is silicon,
Si~\cite{ref19,ref47,ref66}. Spin-orbit effects producing spin
relaxation are much smaller in Si than in GaAs owing to the lower
atomic mass and the inversion symmetry of the crystal structure
maintaining spin-degenerate bands. Furthermore, the most abundant
isotope ${ }^{28}$Si has no nuclear spin, suppressing hyperfine
interactions. These properties make relatively long spin lifetimes
in Si.

The studied SiO${ }_2$(Co)/GaAs heterostructures contain quantum
wells formed by the SiO${ }_2$(Co) film due to the difference of the
chemical potentials between the SiO${ }_2$(Co) and the GaAs. Using
another methods to form quantum wells at interfaces in SCs
(molecular beam epitaxy, MOCVD), we can obtain quantum wells with
desired thickness, depth and number of localized electron levels.
Localized levels can be splitted by the exchange interaction with a
FM grown at the interface or by the interaction with a granular film
containing FM nanoparticles. It is need to note that the latter
technology method -- sputtering of the granular film can solve the
problem of the efficient spin injection difficulty due to the
inherent conductivity mismatch between FM metals and
SCs~\cite{ref67}. Variation of the FM nanoparticle concentration
leads to considerable variation in the conductivity of the granular
film and we can reach conductivity correspondence between the FM and
the SC. In some cases, the combination FM metal layer with a
granular film with FM nanoparticles can be used. In order to reach
efficient spin polarization of injected electrons, we must increase
energies $\varepsilon_{\lambda }^{{\rm (ex)}}$ (\ref{eq6}) of
exchange-splitted levels and the height of the spin-dependent
potential barrier $W$ (\ref{eq9}). The spin-polarized current is the
sum of electrons surmounting the barrier and electrons tunneling
from exchange-splitted states in the quantum well (Figure
\ref{Fig13}). Manipulation of the tunneling (increase of the
tunneling transparency from the highest sublevel and suppression of
the tunneling from other sublevels) can be realized by extending the
region $[\delta_1,\delta_2]$ and decreasing the potential energy in
the region $[\gamma_1,\gamma_2]$.

\begin{figure*}
\begin{center}
\includegraphics*[scale=.6]{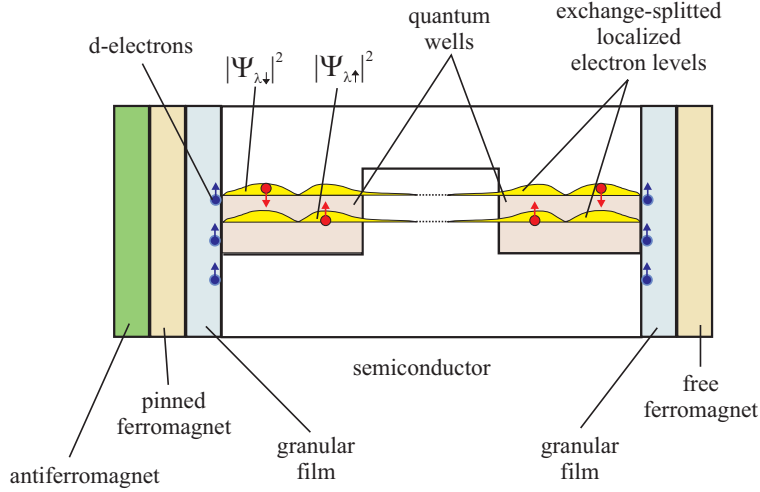}
\end{center}
\caption{Schematic band diagram of the spin-valve structure on the
base of the heterostructure with two granular films and two quantum
wells.} \label{Fig16}
\end{figure*}

Magnetic sensors and non-volatile magnetic memory storage cells can
be constructed on the base of the spin-valve structure containing SC
with two quantum wells and two FM layers, for example, Fe, Co, CoFe
and NiFe alloys (figure \ref{Fig16}). In order to overcome the
conductivity mismatch, the combination FM metal layer with a
granular film with FM nanoparticles sputtered on the SC interface
can be used. In the spin-valve structure one of the FM layers is
exchanged biased using an antiferromagnetic layer (for example,
IrMn, Mn, Ru) and the second is free. The relative magnetizations of
these FM layers can be modulated by manipulating an external
field~\cite{ref46,ref68,ref69}. The FM layers have different
magnetic coercivities to obtain parallel and anti-parallel
alignment. If magnetizations of the layers are aligned parallel to
one another, then spin-polarized electrons easy tunnel from
sublevels of the first quantum well to sublevels of the second one.
In the opposite case, if magnetizations of the layers are
antialigned, then spin polarizations of these sublevels have
anti-parallel alignment. This leads to sharp decreasing in the
tunneling transparency. Thus, the manipulation of magnetic
polarizations gives rise to an efficient spin-filter effect.

\section{Conclusion}

We have studied the electron spin transport in SiO${ }_2$(Co)/GaAs
and SiO${ }_2$(Co)/Si heterostructures, where the SiO${ }_2$(Co)
structure is the granular SiO${ }_2$ film with Co nanoparticles and
have obtained the following results.

\noindent (1) In SiO${ }_2$(Co)/GaAs heterostructures the giant
injection magnetoresistance (IMR) effect is observed. The IMR effect
has positive values and the temperature-peak type character. The
temperature location of the effect depends on the Co concentration
and can be shifted by the applied electrical field. For the SiO${
}_2$(Co)/GaAs heterostructure with 71 at.\% Co the IMR value reaches
1000 ($10^5$ \%) at room temperature, which is two-three orders
higher than maximum values of GMR in metal magnetic multilayers and
TMR in magnetic tunnel junctions. On the contrary, for SiO${
}_2$(Co)/Si heterostructures magnetoresistance values are very small
($4 \%$) and for SiO${ }_2$(Co) films the intrinsic
magnetoresistance has an opposite value.

\noindent (2) High values of the magnetoresistance effect in SiO${
}_2$(Co)/GaAs heterostructures have been explained by
magnetic-field-controlled process of impact ionization in the
vicinity of the spin-dependent potential barrier formed in the
accumulation electron layer in the semiconductor near the interface.
Kinetic energy of electrons, which pass through the barrier and
trigger the avalanche process, is reduced by the applied magnetic
field. This electron energy suppression postpones the onset of the
impact ionization to higher electric fields and results in the giant
magnetoresistance. Although it is need a detailed theoretical model
to understand the effect, the developed model can explain some
features of experimental results. The spin-dependent potential
barrier is due to the exchange interaction between electrons in the
accumulation electron layer in the SC and $d$-electrons of Co.
Existence of localized electron states in the accumulation layer
results in the temperature-peak type character of the barrier and
the IMR in the SiO${ }_2$(Co)/GaAs. The temperature-peak type
character distinguishes the spin-dependent potential barrier from
the Schottky barrier. Maxima of peaks correspond to one
exchange-splitted level in the accumulation layer. The temperature
peak width is inversely proportional to the surface probability of
the Co particle distribution at the interface. In contrast, for
SiO${ }_2$(Co)/Si heterostructures the accumulation layer has small
thickness without any localized states, is tunnel transparent and
does not influence on the injection current.

\noindent (3) FM/SC heterostructures with quantum wells with
spin-polarized localized electrons in the SC at the interface are
proposed as efficient room-temperature spin injectors and magnetic
sensors.

\section*{Acknowledgment}
The authors gratefully acknowledge the assistance of Dr. V.M.
Lebedev (PNPI, Gatchina, Leningrad region, Russia) for determination
of the film composition, Dr. M.V. Baidakova (A.F. Ioffe
Physico-Technical Institute, St. Petersburg, Russia ) for the
measurements by the small-angle X-ray scattering method, and S.Yu.
Krasnoborod'ko (NT-MDT, Russia) for MFM images of the domain
structure of samples. We would like to thank Prof. Yu.G. Kusrayev,
Dr. V.I. Kozub, and Dr. V.L. Korenev for useful discussions. This
work was supported by the Russian Foundation for Basic Research.


\end{document}